\documentclass[intlimits,twoside,a4paper]{article}
\usepackage{graphicx}
\usepackage{amsmath,amssymb}

\usepackage[T2A]{fontenc}
\usepackage[cp1251]{inputenc}

\usepackage{cmpj2}

\newcommand {\lp}{{\rm 1\hskip-0.09cm l}}

\issue{2014}{17}{2}{23002}
\doinumber{10.5488/CMP.17.23002}

\title[On the particle-hole symmetry of the fermionic spinless Hubbard model in $D=1$]%
{On the particle-hole symmetry of the fermionic spinless
Hubbard model in $D=1$}
\author[M.T. Thomaz, E.V. Corr\^ea Silva, O. Rojas]{
M.T. Thomaz\refaddr{AutorTeresa},
E.V. Corr\^ea Silva\refaddr{AutorEdu},
O. Rojas\refaddr{AutorOnofre}}
\addresses{
\addr{AutorTeresa}
	Instituto de F\'{\i}sica, Universidade Federal Fluminense,
	Av. Gal. Milton Tavares de Souza s/n$^{\textit o}$,
	CEP 24210-346, Niter\'oi-RJ, Brazil
\addr{AutorEdu}
	Departamento de Matem\'atica, F\'{\i}sica e Computa\c{c}\~ao, Faculdade
	de Tecnologia, Universidade do Estado do Rio de Janeiro, Rodovia
	Presidente Dutra km 298 s/n$^{\textit o}$, P\'olo Industrial, CEP
	27537-000, Resende-RJ, Brazil
\addr{AutorOnofre}
	Departamento de Ci\^encias Exatas, Universidade Federal de Lavras,
	Caixa Postal 3037, CEP 37200-000, Lavras-MG,  Brazil
}

\date{Received January 10, 2014, in final form February 25, 2014}

\begin{document}

\maketitle

\begin{abstract}
We revisit the particle-hole symmetry
of the one-dimensional ($D=1$) fermionic spinless Hubbard model,
associating that symmetry to the invariance of the
Helmholtz free energy of the one-dimensional spin-$1/2$
$XXZ$ Heisenberg  model, under  reversal of the
longitudinal magnetic field  and at any finite temperature.
Upon comparing two regimes of that chain model
so that the number of particles in one regime
equals the number of holes in the other,
one finds that, in general,  their thermodynamics is
similar, but not identical: both models share
the specific heat and entropy functions, but not the internal energy per
site, the first-neighbor correlation functions, and the number of
particles per site.  Due to that symmetry, the difference
between  the first-neighbor correlation functions
is proportional to the $z$-component of
magnetization of the $XXZ$ Heisenberg model.
The results presented in this paper are valid for any value
of the interaction strength parameter $V$, which describes
the attractive/null/repulsive interaction of neighboring fermions.
\pacs 05.30.Fk, 71.27.+a, 75.10.Pq
\end{abstract}


The one-band Hubbard model \cite{hubbard1,hubbard2} partially describes
quantum magnetic  phenomena; the complexity of real materials,
however, imposes severe limitations on  the direct comparison of
experimental and theoretical results.
It is not always clear which missing terms should be
included in the fermionic Hamiltonian to account for the diversity of
phenomena in a strongly correlated electron system.

The development of optical lattices over the last two decades has
made the experimental simulation of chain models possible.
The three-dimensional Hubbard model at low temperatures has
been simulated by a fermionic quantum gas trapped in an optical
lattice \cite{jordens,schneider}.
A review of the simulation of the Fermi-Hubbard model
with fermionic atoms in optical lattices  can be found in \cite{esslinger}.
The simulation of a one-dimensional
spin-$1/2$ Ising model by a degenerate Bose gas of rubidium atoms
confined in an optical lattice can be found in \cite{simon}.
The simplest one-dimensional fermionic model is the fermionic
spinless Hubbard model,  the generalizations  of which have been
applied to  the description of Verwey metal-insulator transitions
and charge-ordering phenomena of the
Fe$_3$O$_4$, Ti$_4$O$_7$, LiV$_2$O$_4$ and
other $d$-metal compounds~\cite{verwey,kobayashi}.

In this paper we revisit the consequences of the particle-hole
symmetry on the thermodynamics of the one-dimensional fermionic spinless
Hubbard model in the whole range of temperatures, by mapping
it into the exactly solvable  $D=1$ spin-$1/2$ $XXZ$ Heisenberg
model. Appendix \ref{Apend_A} shows the $\beta$-expansion  of the
Helmholtz free energy (HFE) of  this model,
up to the order $\beta^6$~\cite{EPJB2005}.

The spinless fermionic Hubbard model in $D=1$ is a
very simple anti-commutative model  the Hamiltonian of which
is~\cite{sznajd}:
\begin{eqnarray}  \label{1}
{\bf H} (t, V, \mu) = t \sum_{i=1}^{N} \left({\bf c}_i^{\dag} {\bf c}_{i+1}
  +  {\bf c}_{i+ 1}^{\dag} {\bf c}_{i}\right) + V \sum_{i=1}^{N} {\bf n}_i {\bf n}_{i+1}
      - \mu \sum_{i=1}^{N} {\bf n}_i \, ,
\end{eqnarray}
in which $({\bf c}_i, {\bf c}_i^{\dag})$,  with
$i\in\{ 1, 2, \ldots, N\}$, are fermionic annihilation and creation operators, respectively, and $N$ is the number of
sites in the chain. These operators satisfy
anticommutation relations, $ \{ {\bf c}_i, {\bf c}_j^{\dag} \} = \delta_{i j} \lp_i$
and $\{{\bf c}_i, {\bf c}_j \} = 0$,  in which $t$ is the hopping integral, $V$
is the strength of the repulsion ($V>0$) or attraction ($V<0$) between
first-neighbour fermions, $\mu$ is the chemical potential and
${\bf n}_i = {\bf c}_i^{\dag}  {\bf c}_i$ is the
operator number of fermions at the $i^{\rm th}$ site of the
chain.

Sznajd and Becker \cite{sznajd} have shown that the Hamiltonian (\ref{1})
has a particle-hole symmetry; consequently,
the HFE of this model, $W (t, V, \mu; \beta)$,  satisfies the
relation
\begin{eqnarray}  \label{2}
  W (t, V, \mu; \beta) = W (t, V, -\mu + 2V; \beta) - (\mu -V),
\end{eqnarray}
in which $\beta = \frac{1}{k T}$,  $k$ is the Boltzmann's constant
and $T$ is the absolute temperature in Kelvin. The relation (\ref{2}) is valid
for any values (positive, null or negative) of $V$ and $\mu$.
Equation (\ref{2}) provides the condition for having the same number
of particles  and holes at the same potential $V$ but at distinct
chemical potentials,
\begin{eqnarray} \label{3}
 \langle {\bf n}_i \rangle (t, V, \mu; T) = 1
          - \langle {\bf n}_i \rangle (t, V, - \mu + 2V; T),
\end{eqnarray}
\noindent in which $\langle {\bf n}_i \rangle$ is the average number of
fermionic particles at each  site of the chain at temperature $T$.

Haldane \cite{haldane} showed the equivalence
of the model (\ref{1}) and the spin-$1/2$ $XXZ$ Heisenberg model in $D=1$.
More recently, Sznajd and Becker \cite{sznajd} also
used the inverse Wigner-Jordan transformation
to show  that the Hamiltonian (\ref{1}) is mapped onto the Hamiltonian
of the  one-dimensional spin-$1/2$ $XXZ$ Heisenberg model
with a longitudinal magnetic field,
\begin{eqnarray}  \label{4}
{\bf H}_{S = 1/2}(J, \Delta, h) =  \sum_{i=1}^{N} \left[ J \left( {\bf S}_i^x {\bf S}_{i+1}^x
+ {\bf S}_i^y {\bf S}_{i+1}^y  + \Delta {\bf S}_i^z {\bf S}_{i+1}^z  \right)
      - h {\bf S}_i^z\right]  ,
\end{eqnarray}
\noindent in which ${\bf S}^l = \frac{\sigma^l}{2}$,
$l\in\{ x, y, z\}$, and $\sigma^l$  are the Pauli
matrices. The norm of the spin operator
$\vec{\bf S}$ is $||\vec{\bf S}|| = \frac{\sqrt{3}}{2}$.

The Hamiltonians (\ref{1}) and (\ref{4}) are related by
\begin{eqnarray}  \label{5}
 {\bf H} (t, V, \mu) = {\bf H}_{S = 1/2}(J = 2t, \Delta = {V}/({2t}), h = \mu - V)
     - N \left( \frac{J \Delta}{4} + \frac{h}{2} \right)  \lp
\end{eqnarray}
\noindent and $\lp$ is the identity operator of the chain. This relation
shows a constant shift between the energy spectrum of
these two models.

Let $W_{S = 1/2} (J, \Delta, h ; \beta)$ be
the HFE associated to the Hamiltonian (\ref{4}) of the $D=1$ spin-$1/2$ $XXZ$
Heisenberg model. A direct consequence of
(\ref{5}) is that
\begin{eqnarray}   \label{6}
W (t, V, \mu; \beta) = W_{S = 1/2}(J = 2t, \Delta = {V}/({2t}), h = \mu - V; \beta)
     + \left( \frac{V}{4} - \frac{\mu}{2}  \right)  .
\end{eqnarray}

At finite temperature  ($T \not= 0$),  the HFE of the one-dimensional
$S=1/2$ $XXZ$ Heisenberg model is an even function of the longitudinal
magnetic field $h$,
\begin{eqnarray}  \label{7}
  W_{S = 1/2}(J, \Delta, - h; T) =  W_{S = 1/2}(J, \Delta,  h; T) .
\end{eqnarray}
\noindent  Such invariance of  $ W_{S = 1/2}$ comes from the  symmetry
of the Hamiltonian (\ref{4})  upon reversal of the external magnetic field,
$h \rightarrow -h$, and of the spin operators,
$\vec{\bf S}_i \rightarrow -\vec{\bf S}_i$, in which
$i \in \{1, 2, \cdots N\}$.

Consider, for a given magnetic field $h$ and a fixed value (positive,
null or negative) of $V$, the chemical potential $\mu$ so that $ h = \mu- V$.
For a reversed magnetic field, the corresponding chemical potential
$\mu_2$ for which $-h = \mu_2 - V$ is
\begin{eqnarray}  \label{8}
   \mu_2 = - \mu + 2V  .
\end{eqnarray}

The identity (\ref{7}) and the condition (\ref{8})
recover  the result (\ref{2}) satisfied by the HFE of the
spinless Hubbard model for any values of $V$ and $\mu$.
Notice that in the half-filling condition ($\mu = V$),
we have $\mu_2 = \mu$, and there is no visible
consequence of the symmetry (\ref{7}).

We point out that the quantity $- \mu + 2V$, which appears on the
r.h.s. of (\ref{8}), also appears as an argument of $W$ (the HFE of $D=1$
spinless fermionic Hubbard model) in the r.h.s. of (\ref{2}), which in
its turn comes from the particle-hole symmetry of the Hamiltonian
(\ref{1}). On the other hand, (\ref{7}) comes from the fact that the HFE
of the $D=1$ spin-$1/2$ $XXZ$ Heisenberg model is insensitive to a
reversal of the longitudinal magnetic field.

Equation (\ref{3}) can be interpreted as follows: the number of {\it
particles} in the chain under a potential $V$ and a chemical potential
$\mu$ equals the number of {\it holes} in the chain under the same
potential $V$ and a chemical potential $\mu_2$ given by (\ref{8}).
Those configurations correspond to distinct distributions of the fermionic
particles in the chain, and certainly have
some different thermodynamic functions at temperature $T$.
In what follows, we  explore the
consequences of the equality (\ref{7}) in the thermodynamic
functions  of  those two configurations.

The specific heat $C$ and the entropy $S$, both per site, are related
to the HFE of the model (\ref{1}) by
$C(t, V, \mu; \beta) = - \beta^2 \frac{\partial^2 }{\partial \beta^2}
{\left[\beta\ W(t, V, \mu; \beta)\right]}$
and $\frac{S}{k}  = \beta^2 \frac{\partial}{\partial \beta} W(t, V, \mu; \beta),$
respectively. Due to equation (\ref{6}) we obtain
\begin{subequations}
\begin{eqnarray} \label{9a}
C (t, V, \mu; T) =  C (t, V, -\mu + 2V; T)
\end{eqnarray}
\noindent and
\begin{eqnarray}  \label{9b}
 S (t, V, \mu; T) =  S (t, V, -\mu + 2V; T) .
\end{eqnarray}
\end{subequations}
\noindent 	Both (\ref{9a}) and (\ref{9b}) are valid
in the whole range of temperatures $T>0$. This can be verified
at each order  of the $\beta$-expansion of the thermodynamic
functions derived from the expansion (\ref{A.1}) of the HFE of the
model, shown in appendix A.

However, not all thermodynamic functions
of the model (\ref{1}) are identical for the chemical
potentials $\mu$ and $\mu_2$, at the same potential $V$.  The internal
energy per site
$\varepsilon (t, V, \mu; \beta)
= \frac{\partial}{\partial  \beta}[\beta W(t, V, \mu; \beta)]$
distinguishes the distributions of the fermionic particles in the chain:
\begin{eqnarray}\label{10}
\varepsilon (t, V, \mu_2; \beta)  = - V + \mu + \varepsilon (t, V, \mu; \beta).
\end{eqnarray}
\noindent Notice that the difference of internal energies per site
does not depend on the temperature. This equality
is valid for any temperature $T>0$ and it is
verified at each order of the expansion in $\beta$ for this thermodynamic
function, obtained from (\ref{A.1}).

In the spin-$1/2$ Heisenberg model (\ref{4}),
the parallel and anti-parallel configurations of spin with respect
to the external magnetic field can be distinguished, for instance,
by the average value of the $z$-component of the spin operator
${\bf S}_i^z$ at a site and the correlation function of odd powers of such operators.
In terms of fermionic operators, we have
${\bf S}_i^z  = {\bf n}_i - \frac{1}{2}\lp_i$,  in which
$\lp_i$ is the identity operator at the  $i$-th site.

For the spinless fermionic Hubbard model,
the first-neighbor correlation function
${\bf G}_{i, i+1}(t, V,\mu; T) \equiv \langle {\bf n}_i {\bf n}_{i+1}\rangle $
also relates configurations in which the number of particles in one
equals the number of holes in the other,
for two values of the chemical potential, $\mu$ and $\mu_2$.

The two-point correlation  function ${\bf G}_{i, i+1}$ is related to the HFE by
\begin{eqnarray}  \label{11}
 {\bf G}_{i, i+1} (t, V, \mu; T) = \frac{\partial W(t, V, \mu; T)}{\partial V} \,.
\end{eqnarray}

From relation (\ref{6}), the symmetry  condition (\ref{7})
and the definition of the $z$-component of the magnetization of the
$D=1$ spin-$1/2$ $XXZ$ Heisenberg model,
\begin{eqnarray} \label{12}
M_z^{S=1/2} (J, \Delta, h; T)
    &=&  - \frac{\partial W_{S= 1/2}(J, \Delta, h; T) }{\partial h}
         \nonumber   \\
    & = & \langle {\bf S}_i^z \rangle (J, \Delta, h; T),
\end{eqnarray}
\noindent  in which $i \in \{ 1, 2, \cdots, N\}$
and $\langle {\bf S}_i^z \rangle (J, \Delta, h; T)$
is the mean value of the $z$-component of the spin-$1/2$ operator at
the $i^{\rm th}$ site of the chain and  at temperature $T$,  we obtain
\begin{eqnarray}  \label{13}
 {\bf G}_{i, i+1} (t, V, \mu_2; T) -  {\bf G}_{i, i+1} (t, V, \mu; T)
     = - 2 M_z^{S=1/2} (J, \Delta, h; T),
\end{eqnarray}
\noindent where on its r.h.s. we have the $z$-component
of the magnetization $M_z^{S=1/2}$ in the presence of a longitudinal magnetic
field. (Notice that $M_z$ is a one-point function of the model, whereas
${\bf G}_{i,i+1}$ is a two-point function.)  Equation (\ref{13}) is valid
for each order of the $\beta$-expansion of the function
${\bf G}_{i, i+1} (t, V, \mu; \beta)$, derived from
the expansion (\ref{A.1}).

As a consequence of the  symmetry in equation (\ref{7}), the magnetization
$M_z^{S=1/2}$ is an odd function of the magnetic field $h$,
\begin{eqnarray}  \label{14}
M_z^{S=1/2} (J, \Delta, -h; T) = - M_z^{S=1/2} (J, \Delta, h; T).
\end{eqnarray}
\noindent By writing equation (\ref{14}) in terms of fermionic operators,
${\bf S}_i^z  = {\bf n}_i - \frac{\lp_i}{2}$,  one obtains
\begin{eqnarray} \label{15}
 \langle {\bf n}_i \rangle (t, V, \mu; T) = 1
          - \langle {\bf n}_i \rangle (t, V, - \mu + 2V; T),
\end{eqnarray}
\noindent  thus, recovering equation (\ref{3}).

In summary, we have verified that the particle-hole symmetry of the
one-dimensional spinless fermionic Hubbard model (\ref{1})
is associated to the invariance of the HFE of the
$D=1$ spin-$1/2$ $XXZ$ Heisenberg model (\ref{4})
with respect to a reversal of the longitudinal external magnetic field.

The thermodynamics of the chain off the half-filling
condition ($\mu \not= V$) with chemical potentials $\mu$ and $\mu_2$,
under the same potential $V$, are  not  identical; rather, some
thermodynamic functions permit their distinction.
Although the number of fermionic particles in the chain
differ for $\mu$ and $\mu_2$, we obtain unexpected
results, expressed in (\ref{9a}) and (\ref{9b}),
where both configurations exhibit the
same specific heat and entropy per site at any finite temperature
$T$ and at any value of $V$. Distinction arises from other
thermodynamic functions of the chain, though:
the values of the internal energy per site of
these two distributions of particles in the chain
differ by a constant that  is independent of the  temperature;
and  the difference of their first-neighbour correlation functions
is a one-point function proportional to the $z$-component of magnetization
per site of the spin-$1/2$  model (\ref{4}).

The equality of the number of particles in the chain for the chemical
potential $\mu$ and the number of holes in the chain for the
chemical potential $\mu_2$ is a consequence of the odd parity of
magnetization $M_z^{S=1/2} (J, \Delta, h; \beta)$ under reversal of the
magnetic field $h \rightarrow -h$, for any temperature.

The results presented here are valid for any value of $V$ (negative,
null or positive) and any value of  temperature $T>0$, verified at
each order of the $\beta$-expansion of the respective thermodynamic
function. These results are also valid at very low
temperatures and  could be checked in an optical lattice simulation
of the one-dimensional fermionic spinless Hubbard model.

\section*{Acknowledgements}

E.V. Corr\^ea Silva thanks CNPq (Fellowship CNPq, Brazil, Proc.
No.~303876/2010-7) for partial financial support. O. Rojas thanks CNPq and
FAPEMIG for partial financial support.



\appendix

\section{The HFE of the one-dimensional spinless fermionic Hubbard
model up to order $\beta^6$}  \label{Apend_A}

In reference \cite{EPJB2005}  we calculated the $\beta$-expansion
of the HFE of the normalized one-dimensional spin-$S$ of the
$XXZ$ Heisenberg model with single-ion anisotropy term in the
presence of a longitudinal magnetic field  up
to the order $\beta^6$, in terms of the rescaled spin operator
$\vec{s} = {\vec{S}}/{\sqrt{S (S+1)}}$.

In the present work we have applied equation (\ref{6}) to equation (B) of
reference \cite{EPJB2005}, with $||\vec{\bf S}|| = \frac{\sqrt{3}}{2}$, to
derive the $\beta$-expansion, up to the order $\beta^6$, of the HFE of the
one-dimensional fermionic spinless Hubbard model. We have obtained
\begin{eqnarray}
W(t, V, \mu; \beta) & =& -\frac{\ln2}{\beta} -\frac{1}{2} \mu+ \frac{1}{4} V \nonumber  \\
&& +\left(-\frac{1}{4} t^2 +\frac{1}{4} V \mu
 - \frac{5}{32} V^2 -\frac{1}{8} \mu^2 \right) \beta     \nonumber \\
&& + \left( \frac{1}{16} V \mu^2 -\frac{1}{16} t^2 V +\frac{1}{16} V^3 -\frac{1}{8} V^2 \mu \right) \beta^2
                 \nonumber 
\end{eqnarray}
\begin{eqnarray}
&& + \left( -\frac{1}{48} V  \mu^3 +\frac{1}{16} t^2  \mu^2 + \frac{7}{96} t^2 V^2 + \frac{1}{64} V^2  \mu^2
+\frac{1}{96} V^3  \mu     \right. \nonumber \\
 & & \hspace{4.3mm} - \left. \frac{31}{3072} V^4 + \frac{1}{32} t^4 -\frac{1}{8} t^2 V  \mu +\frac{1}{192}  \mu^4  \right) \beta^3
        \nonumber   \\
&& +  \left( -\frac{7}{256} t^2 V^3 - \frac{23}{384} V^3  \mu^2 - \frac{1}{128} V^5 -\frac{1}{32} t^2 V  \mu^2
+\frac{1}{32} t^4 V   \right. \nonumber   \\
& & \hspace{4.3mm}+ \left. \frac{1}{24} V^2  \mu^3 +\frac{1}{16} t^2 V^2  \mu + \frac{7}{192} V^4  \mu
        - \frac{1}{96} V  \mu^4 \right) \beta^4   \nonumber  \\
&& +\left( -\frac{47}{1536} t^4 V^2 - \frac{1}{96} t^2  \mu^4 -\frac{1}{32} t^4  \mu^2 -\frac{239}{7680} V^5  \mu
     \right.     \nonumber \\
& & \hspace{4.3mm}- \frac{1}{144} t^6  + \frac{287}{36864} V^6 - \frac{31}{1152} V^3  \mu^3 - \frac{1}{2880}  \mu^6
        \nonumber \\
& & \hspace{4.3mm}- \frac{21}{2560} t^2 V^4  +\frac{1}{480} V  \mu^5 + \frac{7}{192} t^2 V^3  \mu + \frac{5}{1536} V^2  \mu^4
        \nonumber \\
& & \hspace{4.3mm}-\left. \frac{23}{384} t^2 V^2  \mu^2 +  \frac{1}{24} V t^2  \mu^3 + \frac{139}{3072} V^4  \mu^2
	+\frac{1}{16} V t^4  \mu\right)  \beta^5   \nonumber \\
&& +\left( \frac{13}{768} V t^2  \mu^4 -\frac{1}{64} t^4 V^2  \mu + \frac{157}{1536} t^2 V^3  \mu^2
 - \frac{29}{10240} V^7   \nonumber  \right. \\
& & \hspace{4.3mm}+  \frac{1}{128} V t^4  \mu^2 - \frac{13}{192} t^2 V^2  \mu^3 - \frac{7}{576} V^4  \mu^3
 - \frac{53}{768} t^2 V^4  \mu    \nonumber \\
& & \hspace{4.3mm}+ \frac{17}{11520} V  \mu^6 + \frac{83}{23040} t^4 V^3
+ \frac{1603}{92160} t^2 V^5 - \frac{17}{1920} V^2  \mu^5    \nonumber \\
& & \hspace{4.3mm}-  \frac{119}{30720} V^5  \mu^2 + \frac{389}{46080} V^6  \mu
	- \frac{11}{768} V t^6    \nonumber \\
& & \hspace{4.3mm}+  \left. \frac{41}{2304} V^3  \mu^4 \right) \beta^6 +O(\beta^7).
\label{A.1}
\end{eqnarray}




\clearpage

\ukrainianpart

\title[]{До симетрії частинка-дірка ферміонної безспінової моделі Габбарда в $D=1$}

\author{
M.T. Томас\refaddr{AutorTeresa},
Е.В. Корреа Сільва\refaddr{AutorEdu},
О. Рохас\refaddr{AutorOnofre}}
\addresses{
\addr{AutorTeresa}
Інститут фізики, Федеральний університет Флуміненсе, Нітерой-РЖ, Бразилія
\addr{AutorEdu}
Факультет математики, фізики та інформатики, технологічний факультет, Державний університет  Ріо-де-Жанейро, Ресенде-РЖ, Бразилія
\addr{AutorOnofre}
Факультет точних наук, Федеральний університет м. Лаврас, Лаврас-МЖ, Бразилія
}

\makeukrtitle

\begin{abstract}
\tolerance=3000%


Ми наново переглядаємо симетрію частинка-дірка одновимірної ($D=1$) ферміонної
безспінової моделі Габбарда, пов’язуючи цю симетрію з інваріантністю вільної
енергії Гельмгольца одновимірної спiн-1/2 $XXZ$ моделi Гайзенберга, при
інверсії (перекиданні) поздовжнього магнітного поля і при довільній скінченній
температурі. В результаті порівняння двох режимів ланцюжкової моделі, коли
число частинок в одному режимі дорівнює числу дірок в іншому, знайдено, що в
загальному, їх термодинаміка є подібною, але не ідентичною: обидві моделі
мають однакові функції питомої теплоємності та ентропії, але різні внутрішню
енергію на вузол, кореляційні функції перших сусідів і число частинок на
вузол. Завдяки цій симетрії, різниця між кореляційними функціями перших
сусідів є пропорційною до $z$-компоненти намагніченості $XXZ$ моделі
Гайзенберга. Представлені в цій статті результати справедливі для довільного
значення параметра сили взаємодії $V$, який описує
притягальну/нульову/відштовхувальну взаємодію сусідніх ферміонів.

\keywords квантово-статистична механіка, сильноскорельована електронна
 система, моделі спінових ланцюжків

\end{abstract}

\end{document}